# An Accurate Geometric Distance to the Compact Binary SS Cygni Vindicates Accretion Disc Theory

J. C. A. Miller-Jones[1]*, G. R. Sivakoff[2,3], C. Knigge[4], E. G. Körding[5], M. Templeton[6], E. O. Waagen[6]

[1]International Centre for Radio Astronomy Research, Curtin University, GPO Box U1987, Perth, WA 6845, Australia.

[2]Department of Physics, University of Alberta, CCIS 4-183 Edmonton, AB T6G 2E1, Canada.

[3]Department of Astronomy, University of Virginia, PO Box 400325, Charlottesville, VA 22904, USA.

[4]School of Physics and Astronomy, University of Southampton, Highfield, Southampton SO17 1BJ, UK.

[5]Department of Astrophysics/IMAPP, Radboud University Nijmegen, PO Box 9010, 6500 GL Nijmegen, the Netherlands.

[6]American Association of Variable Star Observers, 49 Bay State Road, Cambridge, MA 02138, USA.

*To whom correspondence should be addressed. E-mail: james.miller-jones@curtin.edu.au.

**Dwarf novae are white dwarfs accreting matter from a nearby red dwarf companion. Their regular outbursts are explained by a thermal-viscous instability in the accretion disc, described by the disc instability model that has since been successfully extended to other accreting systems. However, the prototypical dwarf nova, SS Cygni, presents a major challenge to our understanding of accretion disc theory. At the distance of 159±12 pc measured by the Hubble Space Telescope, it is too luminous to be undergoing the observed regular outbursts. Using very long baseline interferometric radio observations, we report an accurate, model-independent distance to SS Cygni that places the source significantly closer at 114±2 pc. This reconciles the source behavior with our understanding of accretion disc theory in accreting compact objects.**

The accretion of matter onto a central compact object via a disc powers systems as diverse as active galactic nuclei (AGN), X-ray binaries (XRBs), cataclysmic variables (CVs) and young stellar objects. This universal process directly impacts the surrounding environment by driving powerful outflows and jets. CVs, comprised of a white dwarf accreting via Roche lobe overflow from a low-mass stellar companion, provide excellent laboratories for studying disc accretion. There is a large population of nearby CVs, and during outburst the accretion disc both dominates the observed emission at all wavelengths and exhibits structural changes on accessible timescales.

When the mass transfer rate through an accretion disc is sufficiently high, the disc will remain in a fully-ionized, stable state. At lower accretion rates, the resulting lower temperatures allow the recombination of hydrogen, leading to large changes in the disc opacity. This alters the disc cooling mechanism and causes the disc to become thermally unstable, leading to the semi-regular outburst events seen in dwarf nova CVs. The disc instability model (DIM) describes the resulting behavior of the disc as it undergoes a limit cycle, making repeated transitions between a cool quiescent state and a hot, viscous outburst state. With various modifications, the DIM has been successful in both explaining the outbursts of dwarf novae and providing a basic framework for understanding the outbursts of their more massive analogues, the XRBs (*1*).

SS Cygni is the brightest and best-studied dwarf nova, with an optical light curve containing close to 500,000 observations stretching back to its discovery in 1896. It shows repeated outbursts with a mean recurrence time of 49±15 days (*2*). For a distance of ~100 pc, the outburst properties can be well reproduced by the DIM when including the effects of a truncated inner disc and small variations in the mass transfer rate from the secondary (*3*). However, the optical parallax measured by the Hubble Space Telescope (HST) (*4-6*) places the source at a distance of 159±12 pc, implying it has the highest absolute visual magnitude of any dwarf nova (*7*). The mass accretion rate required during outburst to generate this optical luminosity is then too high to be compatible with the DIM because the mean mass transfer rate becomes high enough that the disc should be persistently in the hot, ionized, stable state. SS Cygni should then appear as a nova-like CV with no outbursts. At a distance of 159 pc, the difference in behavior between SS Cygni and the more stable nova-like CVs with similar binary parameters cannot then be ascribed to a difference in the mean mass transfer rate (*8*). This would contradict not only the key prediction of the DIM, but our entire understanding of accretion discs in CVs (*9*) and other accreting systems (*10,11*). The only possible explanation would be an enhancement in mass transfer rate during outburst (*12*), but the required enhancement factor of 150 is implausibly high (*13*).

Since SS Cygni is known to emit radio emission during outbursts (*14*), we used Very Long Baseline Interferometry (VLBI) to measure a model-independent distance based on trigonometry alone. We observed the source with the Very Long Baseline Array (VLBA) and the European VLBI Network (EVN), over 10 epochs between 2010 April and 2012 October (*15*). All observations were triggered following the optical detection of an outburst by the American Association of Variable Star Observers (AAVSO). Fitting the measured positions for a reference position, proper motion and parallax (Table 1), we find a best-fitting distance to SS Cygni of 114±2 pc, with a reduced $\chi^2$ value of 1.1 (Figure 1). The HST distance of 159 pc is ruled out at a very high level of significance (a reduced $\chi^2$ value of 59).

The origin of this discrepancy is unclear. One key advantage of radio parallaxes lies in their use of extragalactic calibrator sources, allowing a direct measurement of the target parallax. By contrast, optical measurements are made relative to nearby stars, so that the mean parallax of the reference stars has to be added to the measured parallax of the target. Therefore, the discrepancy could in principle be due to a problem with one or more of the optical reference stars. However, the difference between our radio parallax and the uncorrected HST optical parallax is more than twice the mean optical reference parallax of 2.23±0.22 milli-arc sec, and larger than the parallax of every single optical reference star *(4,5)*. Thus, to reconcile the radio and optical measurements, there would have to be a very large error in one or more of the reference star parallaxes.

Another possible resolution relates to an inherent bias affecting parallax measurements *(16)*. Measurement errors imply that sources at both larger and smaller intrinsic distances could be measured to have a given parallax. However, for a uniform density of sources, the increased volume of a larger spherical shell makes it more probable that for a given parallax measurement the real source distance is larger, and so measured parallaxes tend to be biased high. Although the HST measurement was corrected for the expected bias of 1.6% *(5)*, the uncertainty on such corrections can be quite large *(17)*. The precision of our radio parallax makes it largely immune to this bias.

Our measured proper motion (117.3±0.2 milli-arc sec year$^{-1}$ along a position angle of 73.5° east of north) agrees with the value given in the fourth version of the US Naval Observatory CCD Astrograph Catalog (UCAC4) *(18)*, but is inconsistent with that measured by *(5)*, who find a similar magnitude but a very different position angle. Together with the systemic radial velocity *(19)*, our fitted distance and proper motion imply a peculiar velocity of 55 km s$^{-1}$ relative to the Galactic rotation; higher than expected for long period CVs *(20)*, but still lower than the 81 km s$^{-1}$ implied by a distance of 159 pc.

Our best fitting distance of 114±2 pc is in excellent agreement with the distance of ~117 pc calculated by *(7)* for SS Cygni to be compatible with the DIM. As shown in Figure 2, it brings the mean mass transfer rate down to 2.6x10$^{17}$ g s$^{-1}$ *(7)*, well below the critical mass transfer rate of 9.0x10$^{17}$ g s$^{-1}$ *(8)*. Thus the disc will be unstable, and the DIM can successfully explain the observed semi-regular outbursts. Our fitted distance also brings the peak absolute magnitude of SS Cygni back in line with that expected for systems of similar orbital period *(21)*.

Our revised distance also tests a second prediction of the DIM: the mass accretion rate at the onset of the outburst decline should be very close to the critical rate. The mass accretion rate predicted by the DIM was approximated by *(7)* to have a power law dependence on several key system parameters (the distance, inclination angle, disc size, white dwarf mass, and apparent magnitude). For a distance of 114 pc and the range of system parameters derived by *(22)*, this approximation predicts an apparent visual magnitude in the range 8.8-9.5 at the onset of decline, as compared to the observed value of ~8.6. If the DIM is correct, and the approximation of (*7*) is valid in the revised region of parameter space, this implies that the inclination angle of the system must be close to 45º, and the mass of the white dwarf should be close to 1 $M_\odot$.

Besides CVs, the DIM has been used to explain the main features of XRB outbursts, when modified to account for the heating effect of X-ray irradiation (*23*) and the truncation of the inner disc (*24*). The thermal-viscous instability at the heart of the DIM is also believed to be active in AGN (*25*), although in this case it only produces relatively small amplitude variability rather than large outbursts (*26*). In an exact parallel to the distinction between dwarf novae and nova-like CVs, the difference between transient and persistent XRBs is attributed to mass transfer rates either below or above a critical threshold (*10,11*). A failure of this model in SS Cygni would therefore have called into question our understanding not only of dwarf novae, but of also transient XRBs.

Finally, the controversy over the distance to SS Cygni emphasizes the need for accurate distances to accreting objects, and highlights the important role that VLBI studies can play in distance determination for radio-emitting Galactic sources.

**Acknowledgments:** We thank the amateur astronomers of the AAVSO whose optical monitoring of SS Cygni was crucial for triggering our radio observations. We also thank J. Thorstensen and A. Deller for useful discussions. This research was supported by the Australian Research Council's *Discovery Projects* funding scheme (JCAMJ, GRS; project number DP120102393), by a Discovery Grant from the Natural Sciences and Engineering Research Council of Canada (GRS), a Consolidated Grant from the UK Science and Technology Facilities Council to the Southampton Astronomy Group (CK), and a VIDI grant from NWO, the Netherlands Organisation for Scientific Research (EK; grant number 016.123.356). The data reported in this paper are tabulated in the Supplementary Materials. The National Radio Astronomy Observatory is a facility of the National Science Foundation operated under co-operative agreement by Associated Universities, Inc. The European VLBI Network is a joint facility of European, Chinese, South African and other radio astronomy institutes funded by their national research councils. This work made use of the Swinburne University of Technology software correlator, developed as part of the Australian Major National Research Facilities Programme and operated under license.

**Figure 1**: Parallax signature of SS Cygni in right ascension (A) and declination (B). The measured position of the source has been plotted against time, having subtracted the best fitting reference position and proper motion given in Table 1. Red points show VLBA measurements and blue points (at MJD 55799 and MJD 56063) show e-EVN measurements. The dashed line shows the best fitting parallax signal, and the grey dotted line shows the calculated signal for a distance of 159 pc. Insets show zoom-ins of the regions around each point.

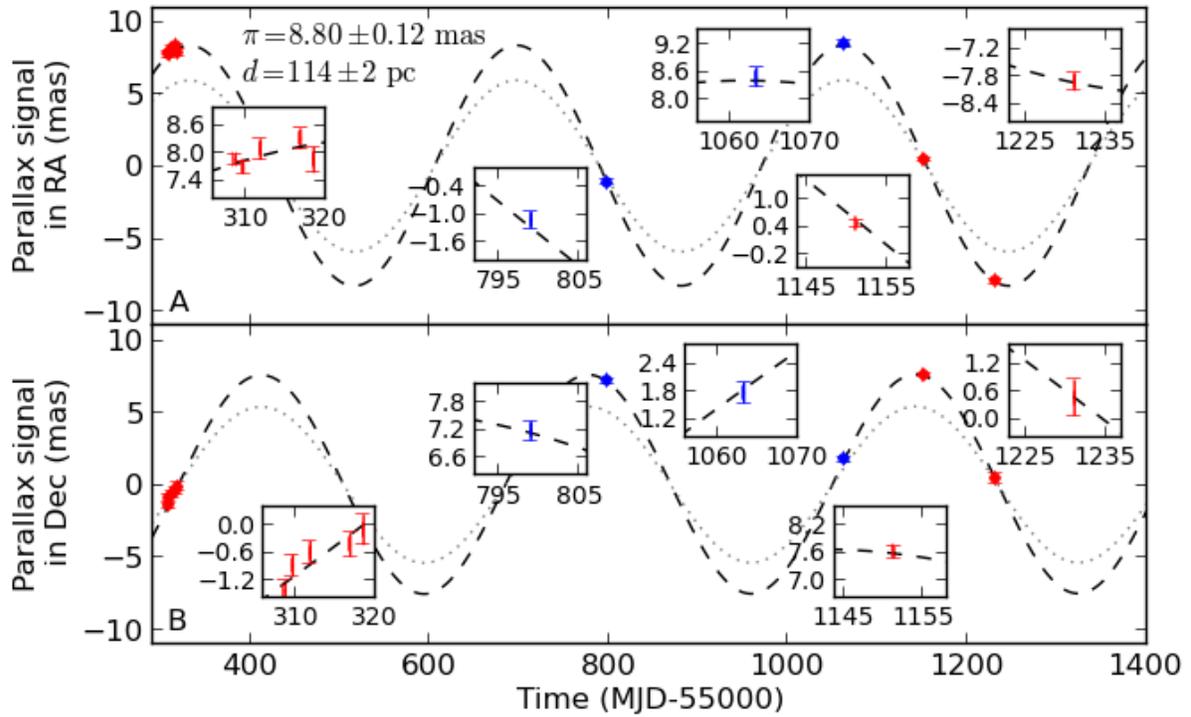

**Figure 2**: Calculated mean mass transfer rates as a function of outer disc radius, as compared to the critical mass transfer rate above which systems are stable against dwarf nova outbursts (black line). Boxes show relatively conservative error regions derived from uncertainties on the system parameters. Values for SS Cygni were calculated from the power-law approximation of (7), using the system parameters tabulated in (8). Values for all other systems were taken directly from (8). Dwarf novae are shown in white, nova-like CVs in light grey, and SS Cygni in dark grey, with the arrow indicating the effect of our distance revision.

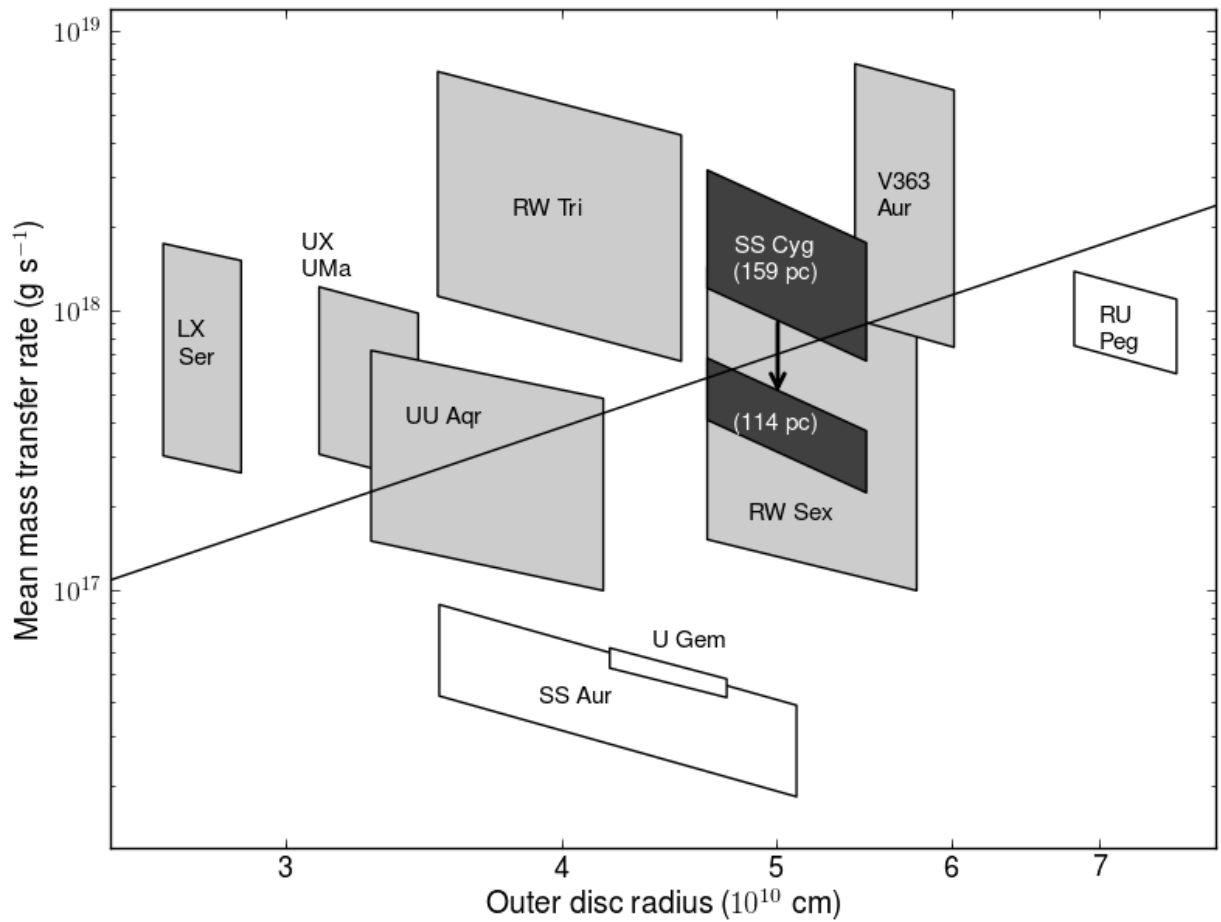

**Table 1:** Fitted astrometric parameters for SS Cyg. Numbers in brackets show the 1σ uncertainty on the last digit.

| Parameter | Fitted value |
|---|---|
| Right ascension $\alpha_0$ (RA; J2000) | 21:42:42.923121(6) |
| Declination $\delta_0$ (Dec; J2000) | 43:35:10.25301(7) |
| Proper motion in RA ($\mu_\alpha \cos \delta$; milli-arc sec year$^{-1}$) | 112.42(7) |
| Proper motion in Dec ($\mu_\delta$; milli-arc sec year$^{-1}$) | 33.38(7) |
| Parallax (milli-arc sec) | 8.80(12) |
| Distance (pc) | 114(2) |
| Reference epoch (MJD) | 55,769.0 |

# Supplementary Materials for

# An Accurate Geometric Distance to the Compact Binary SS Cygni Vindicates Accretion Disc Theory

J. C. A. Miller-Jones\*, G. R. Sivakoff, C. Knigge, E. G. Körding, M. Templeton, E. O. Waagen

\*To whom correspondence should be addressed. E-mail:
james.miller-jones@curtin.edu.au

**This file includes:**

Supplementary Text
Fig. S1
Tables S1 to S2

**Supplementary Text**

**Observations and data analysis**

We observed SS Cygni with Very Long Baseline Interferometry (VLBI) over ten epochs spanning 2010 April through 2012 October. Eight of the epochs were observed at 8.4 GHz with the Very Long Baseline Array (VLBA), and the remaining two with the European VLBI Network (EVN) in electronic VLBI (e-VLBI) mode. The first 6 epochs were taken during a single outburst of SS Cygni, in 2010 April, to provide high angular resolution radio monitoring of the evolution of the outburst. The final four epochs were spaced over the following 2.5 years, and taken to be as close as possible to the peak of the parallax signature in Right Ascension or Declination, given the 49 ± 15 day recurrence time of the outbursts (*2*). All outbursts were triggered on detection of a rising optical flux by the AAVSO. Observations were of duration 6-8 hours, although an archive problem caused the loss of all except the first 2.5 hours of data on 2010 May 2. In most cases the observations were timed such that the midpoint was within an hour of source transit for the central antennas of the array. The exceptions were the VLBA observations of 2010 April 25, 30 and May 6, for which scheduling constraints meant that the observations only began at the time of transit, implying very low source elevations towards the end of these three epochs.

The VLBA observations were made at a central frequency of 8.4 GHz. Data were taken in dual polarization mode, with 64 MHz of bandwidth per polarization. The observations were phase referenced to the nearby calibrator source J2136+4301, at an angular separation of 1.28° from SS Cygni. The phase referencing cycle time was 3 minutes, spending 2 minutes on the target source and 1 minute on the phase reference calibrator in each cycle. Every $7^{th}$ cycle, a scan on the nearby check source J2153+4322 (3.20° from the phase reference source) was substituted for a scan on SS Cygni. Half an hour at the beginning and end of each observation was used to observe bright extragalactic calibrator sources over a wide range of elevations, to correct for unmodelled tropospheric delays and clock errors and hence improve the accuracy of the phase referencing. We used all available VLBA stations, although maintenance issues meant that one or two stations were often missing from the array (see Table S1). The data were correlated using the VLBA-DiFX software correlator (*27*).

The two e-EVN epochs were observed at a central frequency of 5.0 GHz, in dual polarization mode, with 128 MHz of bandwidth per polarization. The available stations differed between epochs, and are listed in Table S1. Data were transferred in real time over the internet to the Joint Institute for VLBI in Europe, where they were correlated using the Mk IV hardware correlator. We used the same phase reference and check sources as for the VLBA observations, although the assumed co-ordinates for the phase reference source differed by 0.009 ms of R.A. and 0.590 mas in Declination from that assumed by the VLBA. Since the target positions were derived relative to that of the phase reference source, this would lead to a spurious positional offset in the EVN data. The measured EVN positions were therefore corrected for this positional shift before fitting for the astrometric parameters.

Data reduction was performed using the Astronomical Image Processing System (AIPS) software package. The VLBA data were corrected for updated Earth Orientation Parameters, digital sampling and parallactic angle effects. Amplitude calibration was carried out using gain curves and system temperature measurements. Fringe fitting was performed on a bright calibrator source, before removing unmodelled tropospheric delays and clock offsets using our geodetic blocks. To reduce the astrometric scatter associated with low elevations, we flagged all data taken at elevations below $23°$. Further bad data were edited out before global fringe fitting on the phase reference calibrator was performed. After applying all the derived corrections, we carried out iterative imaging and self-calibration of the phase reference source to obtain the best possible model for the phase reference calibator. We then derived bandpass corrections and reran the fringe fitting on J2136+4301, using our best image as a model, before applying the final calibration to SS Cygni. The final position of SS Cygni was derived by fitting a point source in the image plane. The only occasion on which SS Cygni appeared to deviate from a point source was 2010 April 25, when the short baselines detected tentative evidence for a discrete jet component, well-separated from the core. At all other epochs, SS Cygni was fully consistent with a point source.

**Data weighting schemes**

Although the VLBA consists of ten identical 25-m dishes, the EVN is a more heterogeneous array, with dish sizes ranging from 25 m (Jodrell Bank Mk-2, Onsala-85) to 100 m (Effelsberg). Therefore the data weights will differ significantly between the different EVN baselines. Although a naturally-weighted image (which is optimized for sensitivity) will have the highest signal-to-noise (S/N), and hence the lowest statistical error on the measured source position, it will be dominated by the systematic errors affecting the most sensitive baselines (Effelsberg-Westerbork in our observations). (*28*) found that this effect caused systematic astrometric offsets larger than the statistical uncertainties when the signal-to-noise ratio was >10. They therefore recommended the use of uniform weighting in such cases, but natural weighting for fainter sources when the error budget was dominated by statistical uncertainties. However, their investigation used the Australian Long Baseline Array (LBA), where the most sensitive baselines (between Parkes, phased ATCA and Tidbinbilla) were all relatively short, in which case uniform weighting would have the effect of downweighting the shorter, more sensitive baselines. In other cases, such as the EVN, alternative schemes for downweighting the most sensitive baselines might be more applicable.

We therefore investigated the effect of using different weighting schemes, allowing the baselines to contribute more equally to the final positional measurement and hence reducing the overall systematic error. We investigated the effect of four additional weighting schemes on both robust and naturally-weighted images; unmodified weights, taking the square or fourth root of the input weights, respectively, and making all data weights identical. The slight positional changes arising from the different weighting schemes caused the fitted parallax to vary over a range of 0.3 mas. Both the chi-squared values of the fits, and a comparison with fits to the VLBA data alone (for which, owing to the homogeneous nature of the array, the additional pre-weighting schemes should have had little effect) suggested that using fourth-root or equal pre-weighting together with naturally-weighted images would give the most accurate results.

**Systematic uncertainties**

At the declination of SS Cygni, the main contribution to systematic astrometric errors comes from unmodelled ionospheric or tropospheric delays (with the former dominating at lower frequencies). The typical systematic uncertainty achievable by both the VLBA and EVN can be estimated, and depends on the angular separation between the target and the phase reference source (*29*). For our observations, this is estimated (in R.A./Dec.) to be 0.044/0.059 mas for the VLBA and 0.047/0.086 mas for the EVN. Our geodetic blocks at the start and end of the observations would have improved the likelihood of reaching this level, but it is nonetheless probable that the final systematic uncertainties would have been somewhat higher than this.

One method of estimating the true systematics is to consider the scatter in the positions of our astrometric check source, J2153+4322, whose weighted standard deviation is 0.105 mas in R.A. and 0.089 mas in Dec.. However, since systematic uncertainties scale linearly with the angular separation between target and calibrator (*29*), we scale these values by a factor 0.4, to get estimated systematic errors of 0.042 and 0.036 mas, respectively. With only two EVN observations, it was not possible to split the dataset up into VLBA and EVN subsets and estimate the systematics for each array separately.

An alternative method to estimate the intra-epoch systematics (*28*) is to split the observations into the 4-8 individual frequency sub-bands in which it was observed, and measure the scatter in the positions measured from the individual sub-bands. However, since our S/N was below 10 for all except three of the observations (see Table S1), the source could not be significantly detected in several of the epochs when this technique was employed. However, this method can form the basis of a Monte Carlo bootstrapping technique (*30*), where each epoch can be split into as many sub-bands as the S/N allows, to form a larger pool of positional measurements. From this pool, nine points (corresponding to the number of observing epochs in which the source was detected) were drawn at random, and fitted to determine the astrometric parameters of the sample. After 10,000 trials, the most compact interval containing 68% of the fit results was used as a measure of twice the overall systematic error. This gave an estimated parallax of 8.96 ± 0.08 mas. However, the disadvantage of this method is that it is dominated by the four brightest epochs (including both EVN observations), so may not be representative of the systematics of the sample as a whole.

Finally, we used an iterative method to determine the systematic uncertainties (*28*). We set the systematic uncertainty in R.A. and Dec. to be equal, and increased it until the reduced $\chi^2$ of the fit was equal to 1.0. We then rotated the uncertainty vector in R.A./Dec. space to minimize the $\chi^2$ value, and then increased the length of the vector again until the reduced $\chi^2$ again reached 1.0. When applied to our final data set (taking the fourth root of the data weights and then applying natural weighting), this suggested systematic errors of 0.108 and 0.091 mas in R.A. and Dec., respectively. These systematic uncertainties were added in quadrature to the statistical uncertainties from each positional measurement before using the singular value decomposition (SVD) algorithm to determine the astrometric parameters of the source, as described by (*31*). The final fitted parallax was 8.80 ± 0.12 mas.

Given the use of different observing frequencies for the two VLBI arrays used, we must address the question of whether there is a systematic offset between the positions measured by the VLBA and EVN, caused, for instance, by a frequency-dependent core-shift in the phase reference calibrator J2136+4301. Images of that source showed it to be dominated by a central, compact component containing over 98% of the flux, with a faint westward extension seen in all epochs (see Figure S1). The lack of bright resolved emission makes a core shift less likely. Furthermore, there was no significant systematic shift in the measured check source positions between the EVN and VLBA epochs. To test for the possibility of a systematic offset between EVN and VLBA positions, we discarded the EVN data from our final sample, and found a fitted parallax of $8.72 \pm 0.20$ mas, fully consistent with the value found for the full dataset.

Finally, a jackknife analysis on the full EVN+VLBA data set gave a parallax of $8.80 \pm 0.11$ mas, confirming that there was no underlying bias in the sample, and hence no systematic shift between EVN and VLBA data. The agreement of the jackknife analysis with the results of the iterative chi-squared analysis described above gives us confidence in our final fitted parallax of $8.80 \pm 0.12$ mas. This parallax corresponds to a distance of $114 \pm 2$ pc for SS Cygni.

**Apparent visual magnitude in quiescence:**

Using our revised distance measurement, we can compare the apparent visual magnitude of the system in quiescence to the expected visual magnitude from a K4.5 V star, modified by the flux from the accretion disk and white dwarf (*22*). The radius of the Roche-lobe filling donor is given by the expression of (*32*), and set only by the well-constrained mass ratio. We use the surface brightness of a K4.5 V star given by (*33*) to determine the donor contribution to the visual flux, and the correction factor derived by (*22*) for the contribution of the disk and white dwarf. For a distance of 114 pc, we then predict an apparent visual magnitude of $11.72 \pm 0.34$ in quiescence, consistent with the observed value of 12.0-12.1 (*22*).

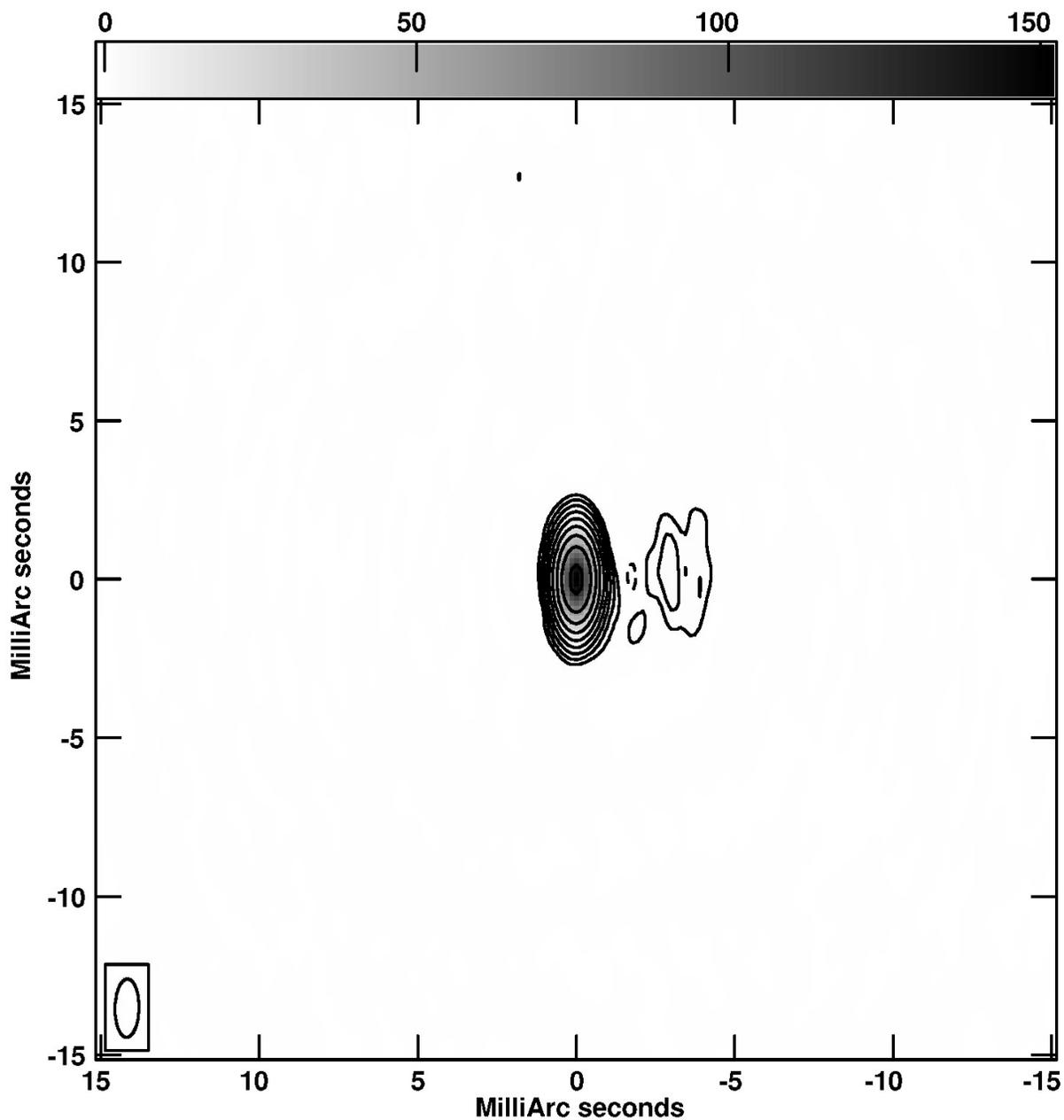

**Fig. S1:** VLBA image of the phase reference calibrator J2136+4301, from 2012 October 30. The rms noise is 0.1 mJy beam$^{-1}$, and the peak brightness is 151.8 mJy beam$^{-1}$. Contours are at levels of ±0.5, 1, 2, 4, 8,… mJy beam$^{-1}$. Greyscale shows the source brightness in units of mJy beam$^{-1}$. The central source dominates the flux density, but there is a faint extension to the west, with an integrated flux density of a few mJy.

**Table S1:** VLBA and e-EVN observations of SS Cygni, showing the Modified Julian Date, exposure time, stations used, and measured source flux density for each epoch.

| Date | MJD | Time on source (min) | Array | Flux density (mJy) |
|---|---|---|---|---|
| 2010-04-22 | 55308.56 ± 0.14 | 148 | VLBA: Br Hn Kp La Mk Nl Ov Pt | 0.47 ± 0.05 |
| 2010-04-23 | 55309.58 ± 0.14 | 127 | VLBA: Br Hn Kp La Mk Nl Ov Pt | 0.39 ± 0.06 |
| 2010-04-25 | 55311.76 ± 0.13 | 130 | VLBA: Br Hn Kp La Mk Nl Ov Pt | 0.29 ± 0.06 |
| 2010-04-30 | 55316.77 ± 0.10 | 157 | VLBA: Br Fd Hn Kp La Mk Nl Ov Pt Sc | 0.39 ± 0.07 |
| 2010-05-02 | 55318.44 ± 0.04 | 62 | VLBA: Br Fd Hn Kp La Mk Nl Ov Pt Sc | 0.25 ± 0.07 |
| 2010-05-06 | 55322.74 ± 0.13 | 206 | VLBA: Br Fd Hn Kp La Mk Nl Ov Pt Sc | <0.15 |
| 2011-08-25 | 55798.95 ± 0.12 | 107 | EVN: Ef Hh Jb Mc On Tr Wb | 3.05 ± 0.11 |
| 2012-05-16 | 56063.17 ± 0.12 | 116 | EVN: Ef Jb Mc Nt Tr Wb Ys | 1.29 ± 0.05 |
| 2012-08-12 | 56151.29 ± 0.12 | 181 | VLBA: Br Fd Hn Kp La Mk Nl Ov Pt Sc | 0.93 ± 0.04 |
| 2012-10-30 | 56231.10 ± 0.11 | 177 | VLBA: Br Fd Kp La Mk Nl Ov Pt Sc | 0.17 ± 0.05 |

**Table S2:** Measured positions for SS Cygni. EVN data have been corrected for the different assumed phase reference calibrator position, as described in the text. Numbers in parentheses show the 1σ statistical uncertainty on the least significant digit.

| MJD | Array | R.A. (J2000) | Dec (J2000) |
| --- | --- | --- | --- |
| 55308.56 ± 0.14 | VLBA | 21 42 42.910803(6) | 43 35 10.2095(1) |
| 55309.58 ± 0.14 | VLBA | 21 42 42.910815(7) | 43 35 10.2101(2) |
| 55311.76 ± 0.13 | VLBA | 21 42 42.91091(2) | 43 35 10.2106(2) |
| 55316.77 ± 0.10 | VLBA | 21 42 42.91108(2) | 43 35 10.2113(2) |
| 55318.44 ± 0.04 | VLBA | 21 42 42.91108(2) | 43 35 10.2117(2) |
| 55798.95 ± 0.12 | EVN | 21 42 42.92387(1) | 43 35 10.2629(1) |
| 56063.17 ± 0.12 | EVN | 21 42 42.93224(1) | 43 35 10.2817(1) |
| 56151.29 ± 0.12 | VLBA | 21 42 42.933995(2) | 43 35 10.29555(4) |
| 56231.10 ± 0.11 | VLBA | 21 42 42.93548(1) | 43 35 10.2957(3) |